\documentclass[reprint,aps,eqsecnum,twocolumn,amsmath,superscriptaddress, amsfonts,amssymb,nofootinbib]{revtex4-2}

\usepackage{comment}

\usepackage{graphicx}
\usepackage{color}
\usepackage{bm}
\usepackage[utf8]{inputenc}
\usepackage[english]{babel}

\newcommand{\beq}{\begin{equation}}
\newcommand{\eeq}{\end{equation}}
\newcommand{\beqs}{\begin{eqnarray}}
\newcommand{\eeqs}{\end{eqnarray}}
\newcommand{\lsim}{\mathrel{\raisebox{-
.6ex}{$\stackrel{\textstyle<}{\sim}$}}}

\newcommand{\tl}{$\tau \to \ell \gamma \gamma$}
\begin{document}


\title{Upper limits on  branching ratios of  the lepton-flavor-violating decays
  $\tau \to \ell \gamma\gamma$ and $\tau \to \ell X$}


\author{Douglas A. Bryman}
\thanks{doug@triumf.ca}
\affiliation{Department of Physics and Astronomy, University of British 
Columbia, Vancouver British Columbia, V6T 1Z1, Canada}
\affiliation{TRIUMF, 4004 Wesbrook Mall, Vancouver, British Columbia V6T
  2A3, Canada}

\author{Shintaro Ito}
\thanks{sito@post.kek.jp}
\affiliation{KEK, 1-1 Oho, Tsukuba, Ibaraki, 305-0801, Japan}

\author{Robert Shrock}
\thanks{robert.shrock@stonybrook.edu}
\affiliation{C. N. Yang Institute for Theoretical Physics and 
Department of Physics and Astronomy, \\
Stony Brook University, Stony Brook, NY 11794, USA }

\begin{abstract}

  From analysis of data produced by the BABAR experiment,   the first  upper bounds (90\% C.L.) were obtained 
  on the branching ratios $Br(\tau \to e \gamma\gamma)<2.5 \times 10^{-4}$ and $Br(\tau \to \mu \gamma\gamma)<5.8 \times 10^{-4}$. In addition, improved upper bounds (95\% C.L.) were found on branching ratios 
  $Br(\tau \to e X  )<1.4 \times 10^{-3}$ and $Br(\tau \to \mu X) <2.0 \times 10^{-3}$,  where  $X$ is an undetected  weakly interacting boson with mass $m_X< 1.6$ GeV/$c^2$.  

\end{abstract}
\maketitle

\section{Introduction}

\label{intro_section}

The violation of lepton family number has been firmly established by
the observation of neutrino oscillations, which also implies charged
lepton family (flavor) number violation (CLFV).  Although no CLFV has been
observed yet, it is of fundamental interest, and searches for CLFV
processes continue to be pursued.  In the Standard Model (SM)
extended to include massive neutrinos (generically denoted the
$\nu$SM), the branching ratios for CLFV decays such as
$\mu \to e \gamma$, 
$\mu \to e \gamma\gamma$, 
$\mu \to ee \bar e$, 
$\tau \to \ell \gamma$, and 
 $\tau \to \ell \ell' \bar \ell'$, where $\ell=e,\mu$ and $\ell'=e,\mu$,
are many orders of magnitude below
the level where they could be observed in existing or planned
experiments.  This means that searches for these decays and similar
CLFV processes are of  great interest as probes of physics
beyond the $\nu$SM (BSM).

Other CLFV processes which are not present in the SM may involve new weakly interacting bosons ($X$). For example, searches for $\mu \to e X$ were  reported in
\cite{bryman_clifford, bilger, jodidio, twist, triumf_muex}. 
The emission of an $X$ boson has also been searched for in $\pi^+$ decays \cite{eichler, triumf88, triumf_pilnx} and $K^+$ decays \cite{yamazaki, baker87, bnl2, na62_klnx}. 
The ARGUS  experiment \cite{albrecht95} at DESY  reported  limits for $Br(\tau \to \ell X)/Br(\tau \to \ell \nu_\tau \bar\nu_{\ell})$.

Current upper bounds \footnote{Unless otherwise indicated, all experimental upper limits
  are given at the 90\% confidence level (C.L.).} on
some CFLV $\tau$ decay modes are listed in Table
\ref{clfv_tau_decay_table}. 
In this paper,  we used existing data to set
the first upper limits on the branching ratios of  the CLFV decays $\tau \to e \gamma\gamma$ and
$\tau \to \mu \gamma\gamma$.  
We also examined the branching ratios for the decays $\tau \to \ell X$.

\begin{table}
  \caption{\footnotesize{Upper limits 
      on the branching ratios for some CLFV decays of the $\tau$ lepton \cite{pdg}.}}
\begin{center}
\begin{tabular}{|c|} \hline\hline
  $Br$(decay) upper limit \\ \hline
$Br(\tau \to e \gamma) < 3.3 \times 10^{-8}$  \\
$Br(\tau \to \mu \gamma) < 4.4 \times 10^{-8}$  \\
$Br(\tau \to e e \bar e) < 2.7 \times 10^{-8}$  \\
$Br(\tau \to e \mu \bar \mu) < 2.7 \times 10^{-8}$  \\
$Br(\tau \to \mu e \bar e) < 1.8 \times 10^{-8}$  \\
$Br(\tau \to \mu \mu \bar \mu) < 2.1 \times 10^{-8}$  \\
$Br(\tau \to e \pi^0) < 0.80 \times 10^{-7}$ \\
$Br(\tau \to \mu \pi^0) < 1.1 \times 10^{-7}$  \\
\hline\hline
\end{tabular}
\end{center}
\label{clfv_tau_decay_table}
\end{table}

\section{Theoretical Background}

\label{theory_motivation_section}

\subsection{CLFV in the $\nu$SM}

To accommodate the observed neutrino oscillations and associated
violation of lepton family number in the neutrino sector, 
the  (renormalizable) SM Lagrangian can be modified by adding a number $n_s$ of
electroweak-singlet neutrino fields $\nu_{i,R}$, $i=1,...,n_s$,
conventionally written as right-handed chiral fermions.
With these,  Yukawa terms are formed  with the left-handed lepton
doublets which, via the vacuum expectation values of the Higgs field,
yield Dirac-type mass terms for neutrinos.  The electroweak-singlet
neutrinos also generically lead to Majorana mass terms of the form $\sum_{i,j}
M^{(R)}_{ij} \nu_{i,R}^T C \nu_{j,R} + h.c.$ The diagonalization of
this combination of Dirac and Majorana mass terms yields the neutrino
mass eigenstates.  The resultant unitary transformation relating the
left-handed chiral components of the mass eigenstates of the
neutrinos, $\nu_{i,L}$, to the weak eigenstates, $\nu_{a,L}$, is given
by
\beq
\nu_{a,L} = \sum_i U^{(\nu)}_{ai} \, \nu_{i,L} \ .
\label{unu}
\eeq
The property that $U^{(\nu)}$ is different from the identity gives
rise to neutrino oscillations and the associated violation of lepton
family number in the neutrino sector.  
(There is, in general, also
violation of total lepton number in the $\nu$SM, due to the presence
of Majorana mass terms.)

The diagonalization of the charged lepton mass matrix involves another
unitary matrix $U^{(\ell)}$, and the product of (the adjoint of)
$U^{(\ell)}$ and $U^{(\nu)}$ determines the form of the weak charged
current:
\beq
J_\lambda = \bar \ell_L \gamma_\lambda \nu_{\ell,L} =
\sum_{a,i} \bar \ell_{a,L} \gamma_\lambda U_{ai}\nu_{i,L} \ ,  
\label{cc}
\eeq
where $U$ is the lepton mixing matrix, 
\beq
U = U^{(\ell) \ \dagger} \, U^{(\nu)} \ . 
\label{u}
\eeq

As an example of CLFV in the $\nu$SM, the branching ratio for $\ell_a \to
\ell_b + \gamma$ is \cite{meg77,ls77}
\beq
Br(\ell_a \to \ell_b \gamma) = \frac{3\alpha_{em}}{32\pi}
\bigg | \sum_i U_{ai}U_{bi}^* \, \frac{m_{\nu_i}^2}{m_W^2} \bigg |^2 \ , 
\label{llg}
\eeq
where $\alpha_{em}=e^2/(4\pi)$ is the fine structure constant, and $a$
is the family or generation index, with $\ell_1\equiv e$, $\ell_2 \equiv \mu$, and
$\ell_3 \equiv \tau$.  Using current data on neutrino masses and lepton
mixing, the resultant SM predictions for the branching ratios for the
decays $\mu \to e \gamma$, $\tau \to e \gamma$, and $\tau \to \mu
\gamma$ have values $\lsim 10^{-53}$, far below a level that could be
observed in any existing or planned experiment.  In passing, we recall
the current upper limits on CLFV muon decays,  $Br(\mu \to e
\gamma) < 4.2 \times 10^{-13}$ from the MEG experiment at PSI
\cite{meg}, $Br(\mu \to e \gamma\gamma) < 0.72 \times 10^{-10}$ from
the Crystal Box experiment at LAMPF \cite{bolton_lampf_megg}, and
$Br(\mu \to e e \bar e) < 1.0 \times 10^{-12}$ from the SINDRUM
experiment at SIN/PSI \cite{sindrum}.  Since the decay $\ell_a \to
\ell_b \gamma \gamma$ involves emission of a second photon, as
compared with $\ell_a \to \ell_b \gamma$, it follows that for the $\nu$SM, 
up to logarithmic terms, 
\beq
Br(\ell_a \to \ell_b \gamma \gamma) \sim
\alpha_{em} \, Br(\ell_a \to \ell_b \gamma).
\label{megg_meg_nusm}
\eeq
(Similarly,  $Br(\ell_a \to \ell_b \ell_c \bar \ell_c) 
\sim \alpha_{em} Br(\ell_a \to \ell_b \gamma)$ in the $\nu$SM.) 

\subsection{Possible Physics Beyond the Standard Model Contributing to $\tau \to \ell \gamma$ and  $\tau \to \ell \gamma \gamma$ }

Although CLFV processes are predicted to be unobservably small in the
$\nu$SM, there are many models of physics beyond the $\nu$SM that
generically predict CLFV at observable rates.  While none of these
models has been confirmed by experiment,  they remain of
interest since they address incomplete aspects of the SM.
One such aspect concerns the Higgs mass. There is a fine-tuning
problem associated with this quantity since one-loop corrections to
the Higgs mass squared are quadratically sensitive to the highest mass
scale in an ultraviolet completion of the SM, such as a grand unified
theory.  Two early ideas for BSM physics that addressed this problem
were supersymmetry (SUSY) and dynamical electroweak symmetry breaking (EWSB),
and both of these generically predicted CLFV (as well as a number of
flavor-changing neutral-current processes) at observable levels. For
example, supersymmetric extensions of the SM predicted the decay $\mu
\to e \gamma$ to occur at observable levels \cite{hall_suzuki,ihl_meg,bhs,hisano2009},
and this is also true of $\tau \to \ell \gamma$. Early SUSY models
with light neutralinos $\tilde \chi$ allowed the decay
$\mu \to e \tilde\chi \tilde\chi$, which would be distinct
from SM $\mu$ decay \cite{barber_shrock}.  Substantial contributions to 
$\mu \to e \gamma\gamma$ and $\tau \to \ell \gamma\gamma$ would also be
expected in such SUSY theories. Although searches for
supersymmetric particles at the Fermilab Tevatron and at the CERN
Large Hadron Collider (LHC) have yielded null results so far, there
still remains the possibility of supersymmetry characterized by a
SUSY-breaking scale that is larger than the electroweak scale.

Dynamical EWSB models also predict CLFV processes at possibly
observable levels \cite{etc,ckm,tc_pdg}. A relevant property of
reasonably ultraviolet-complete dynamical EWSB models is the generic
presence of sequential stages of breaking of an asymptotically
free chiral gauge symmetry in the ultraviolet. 
The feature that the third generation is associated with the lowest 
of these scales could give rise to enhanced CLFV procees involving the
$\tau$ lepton \cite{dml}. Modern versions of dynamical EWSB models typically involve
quasi-conformal behavior, which can result naturally from an
approximate infrared fixed point of the renormalization group
equations describing the strongly coupled vectorial gauge interaction
\cite{wtc,qcconferences,quasiconformal}.  In general, in these models,
the observed Higgs is a composite state.  These dynamical EWSB models
are tightly constrained by precision electroweak data, the observed agreement
of the Higgs boson with SM predictions, and, more generally, the 
non-observation of any BSM Higgs properties at the LHC.  

A large variety of other BSM theories predict CLFV effects at
potentially observable levels. These could have the potential to alter the $\nu$SM 
relation (\ref{megg_meg_nusm}).  
For example, in theories with doubly charged leptons, the ratio of branching ratios 
$Br(\tau \to \ell \gamma\gamma)/Br(\tau \to \ell \gamma)$ 
can be substantially enhanced relative to the $O(\alpha_{em})$ relation in Eq. (\ref{megg_meg_nusm}), just as was true of the ratios  $Br(\mu \to e e \bar e)/Br(\mu \to e \gamma)$ and $Br(\mu \to e\gamma\gamma)/Br(\mu \to e \gamma)$ \cite{wilczek_zee}; some recent studies of theories with doubly charged leptons  \cite{dcl} provide  experimental constraints.  These theories could also lead to an enhancement of 
$\tau \to \ell \pi^0$, which, via the $\pi^0 \to \gamma\gamma$ decay, could contribute to a $\ell\gamma\gamma$ final state and hence to an overall $\tau \to \ell\gamma\gamma$  decay.  

Of particular interest for $\tau \to
\ell X$ decays are models with a light pseudo-Nambu-Goldstone boson
(NGB) or a massless NGB that can couple to fermions in a
flavor-violating manner \cite{wilczek82,grinstein_preskill_wise}. These
arise in models that hypothesize a ``horizontal'' symmetry mixing SM
fermions transforming in the same manner under the SM gauge group,
$G_{\rm SM}={\rm SU}(3)_c \otimes {\rm SU}(2)_L \otimes {\rm U}(1)_Y$,
namely the sets $(e,\mu,\tau)_L$, $(e,\mu,\tau)_R$, and so forth for
the neutrinos and quarks. With the hypothesized generational (i.e.,
family) symmetry taken to be global, a consequence would be that the spontaneous breaking of the flavor symmetry
would lead to massless, spinless NGB(s) (often called familons). In the
presence of some explicit breaking of the generational symmetry, the
spontaneous breaking yields light  NGB(s), with mass(es) determined by the
relative sizes of explicit and spontaneous symmetry breaking.  These  NGBs are often called ``axion-like particles'' (ALPS).  CLFV
effects may also be associated with spontaneous breaking of total
lepton number and resultant majorons
\cite{gelmini_roncadelli,cmp_majoron,gelmini_nussinov_yanagida,berezhiani}.
Some more recent studies and reviews include
\cite{feng, jaeckel_ringwald, jaeckel, heeck_rodejohann, calibbi_review, neubert_alp, cornella, endo_belle2, diluzio, calibbi2020}.

Models featuring extra $Z'$ vector bosons with flavor-non-diagonal
couplings can yield CLFV effects at observable levels (e.g.,
\cite{langacker_zprime,heeck_zprime}). These could contribute to 
CLFV decays such as $\tau \to \ell\gamma\gamma$  and 
$\tau \to \ell\gamma$. CLFV processes have also been
studied in models with extra (spatial) dimensions and fermion fields
having localized wave functions in these extra dimensions
\cite{ng,nuled}. An appeal of these models is that they can produce a
strong hierarchy in SM fermion masses via moderate separation of
fermion wave function centers in the extra dimensions \cite{as,ms}.
Connections between reported anomalies in $B$ meson decays, $e-\mu$
universality violation, and models with CFLV have been discussed in a
number of studies and are reviewed, e.g., in \cite{belle2_physics}.
Although dark matter is, in principle,
independent of CFLV, there may be connections between these in certain
models \cite{snowmass_dm}.

\section{$\tau \to \ell \gamma \gamma$}

\label{tlgg_section}

In the following  we  discuss the angular distribution expected for the decay products 
$\tau \to \ell \gamma\gamma$.  
Then, we use existing data on searches for $\tau \to e \gamma$ and $\tau \to \mu \gamma$  from the BABAR experiment at SLAC \cite{aubert_babar_tlg}  to derive the first upper limits on the branching ratios for these decays, $\tau \to e \gamma \gamma$ and $\tau \to \mu \gamma \gamma$.
(See also the recent result from Belle \cite{belle2021_tmg}, which
improves slightly on the upper limit on $B( \tau \to \mu \gamma)$ in \cite{aubert_babar_tlg}.) 
In abstract
notation, these decays are of the form $\ell_a \to \ell_b \gamma \gamma$ with
the generational indices $a=3$ and $b=1, \ 2$, respectively.

\subsection { Angular Distribution for $\tau \to \ell \gamma\gamma$}

\label{tlgga_section}

A calculation of the decay rate for $\mu \to e \gamma \gamma$ was originally carried out in 1962 by Dreitlein and
Primakoff \cite{dreitlein_primakoff} and was applied to the data on
searches for $\mu \to e \gamma$  to set the upper bound
$Br(\mu \to e\gamma\gamma) < 5 \times 10^{-6}$. This procedure including a general expression for the angular distribution of the photons (discussed below) was
applied by Bowman \emph{et al.} \cite{bclm} to data from two
contemporaneous experiments searching for $\mu \to e \gamma$
\cite{depommier77,povel77} to derive the upper limit $Br(\mu \to e\gamma\gamma) < 5 \times 10^{-8}$  \cite{megg_bounds, azuelos83, bolton88, recentmu, dk_megg}.

For our analysis, we do not assume a particular BSM theory, but instead
use an effective field theory method. In general, for a decay of the
form $\ell_a \to \ell_b \gamma \gamma$, the operators that
contribute to leading order to the effective Lagrangian ${\cal
  L}_{eff,\ell_a\ell_b\gamma\gamma}$ are lepton bilinears contracted
with $F_{\alpha\beta}F^{\alpha\beta}$ or $F_{\alpha\beta} \tilde
F^{\alpha\beta}$ (with coefficients given in Eq. (\ref{leff_tlgg}) below), 
where $F^{\alpha\beta}$ is the electromagnetic field
strength tensor and $\tilde F^{\alpha\beta} = (1/2)
\epsilon^{\alpha\beta\lambda \rho} F_{\lambda \rho}$ is its dual.  At
more suppressed levels, there are additional operators involving
derivatives.  In $d$ spacetime dimensions, the mass dimension of an
operator ${\cal O}$ comprised of a lepton bilinear, a product of $FF$
or $F\tilde F$, and $n_{_\partial}$ derivatives, is ${\rm dim}({\cal
  O}) = 2d-1+n_{_\partial}$.  It follows that the coefficient $c_{\cal
  O}$ has mass dimension ${\rm dim}(c_{\cal O}) =
-(d-1+n_{_\partial})$, i.e., for the physical case $d=4$, ${\rm
  dim}(c_{\cal O}) = -(3+n_{_\partial})$.  One can thus write
\beq
c_{\cal O} = \frac{\bar c_{\cal O}}
                  {(\Lambda_{\cal O})^{3+n_{_\partial}}} \ , 
\label{cogen}
\eeq
where $\bar c_{\cal O}$ is dimensionless and $\Lambda_{\cal O}$
denotes a scale of BSM physics responsible for the appearance of the
operator ${\cal O}$. Since in both of the decays $\tau \to \ell \gamma
\gamma$ with $\ell=e$ or $\ell=\mu$, $m_\ell \ll m_\tau$, the only
mass that enters into the phase space kinematics of the decay is
$m_\tau$.  Because the derivatives yield factors of momenta in the
amplitude, and the sizes of these momenta are set (in the $\tau$ rest
frame) by $m_\tau$, it follows that the contribution of an operator
with $n_{_\partial}$ derivatives is suppressed by the factor
$(m_\tau/\Lambda_{\cal O})^{n_{_\partial}}$. The agreement of
the $\nu$SM with current data implies that the scales $\Lambda_{\cal
  O}$ are much larger than $m_{W,Z}$, and hence $(m_\tau/\Lambda_{\cal
  O})^{n_{_\partial}} \ll 1$.  Therefore, operators with derivatives
are expected to make a negligible contribution to the amplitude for
$\tau \to \ell \gamma\gamma$.  The effective Lagrangian for $\tau \to
\ell \gamma \gamma$ can then be written, retaining non-negligible
terms, as
\begin{widetext}
\beqs
    {\cal L}_{eff,\tau \ell \gamma \gamma } &=&
    c_{\tau \ell \gamma \gamma,LR;FF}[\bar\ell_L \tau_R]
    F_{\alpha\beta}F^{\alpha\beta} +
    c_{\tau \ell \gamma \gamma,RL;FF}[\bar\ell_R \tau_L]
    F_{\alpha\beta}F^{\alpha\beta} +
    c_{\tau \ell \gamma \gamma,LR;F\tilde F}[\bar\ell_L \tau_R]
    F_{\alpha\beta}\tilde F^{\alpha\beta} \cr\cr
    &+&
    c_{\tau \ell \gamma \gamma,RL;F\tilde F}[\bar\ell_R \tau_L]
    F_{\alpha\beta}\tilde F^{\alpha\beta} + h.c. \ ,  
\label{leff_tlgg}
\eeqs
\end{widetext}
where the subscript $LR$ on $c_{\tau \ell \gamma\gamma,LR;FF}$ refers
to the chirality structure in the associated lepton bilinear, $[\bar\ell_L \tau_R]$, and similarly for the other coefficients.
Without loss of generality, we will introduce a single effective mass scale
$\Lambda_{\tau\ell\gamma\gamma}$ to characterize the CFLV physics
responsible for the decay $\tau \to \ell\gamma\gamma$; any differences
in the actual mass scales characterizing different operators ${\cal O}$
are absorbed into the values of the dimensionless coefficients
$\bar c_{\cal O}$.  Then Eq. (\ref{cogen}) reads 
\beq
c_{\cal O} = \frac{\bar c_{\cal O}}{\Lambda_{\tau\ell\gamma\gamma}^3}
\label{cbar}
\eeq
for each of the coefficients $c_{\tau \ell \gamma \gamma,LR;FF}$,
$c_{\tau \ell \gamma \gamma,RL;FF}$, $c_{\tau \ell \gamma
  \gamma,LR;F\tilde F}$, and $c_{\tau \ell \gamma \gamma,RL;F\tilde
  F}$ in ${\cal L}_{eff,\tau \ell \gamma \gamma }$.  Let us denote the
four-momenta of the $\tau$, the final-state charged lepton $\ell$, and
the two photons as $p_\tau$, $p_\ell$, $k_1$, and $k_2$, respectively,
and Lorentz-scalar products of two four-vectors as $p_\tau \cdot
p_\ell$, etc. Let us further denote the matrix element for this decay
as ${\cal M}_{\tau \to \ell\gamma\gamma}$.  As usual, the amplitude is
Bose-symmetrized with respect to the interchange of the identical
bosons (photons) in the final state. With the above input ${\cal
  L}_{eff,\tau \ell \gamma\gamma}$, the square of the amplitude has
a kinematic factor $
(p_\tau \cdot p_\ell)(k_1 \cdot k_2)^2 = 
m_\tau E_\ell [E_{\gamma_1}E_{\gamma_2}(1-\cos\theta_{\gamma_1\gamma_2})]^2$,
and the differential decay rate is 
\begin{widetext}
\beqs
\frac{d\Gamma_{\tau \to \ell\gamma\gamma}}
     {dE_{\gamma_1}dE_{\gamma_2}d\cos\theta_{\gamma\gamma}} \propto
     \bigg( \frac{\sum_{\cal O} |\bar c_{\cal O}|^2}
     {\Lambda_{\tau\ell\gamma\gamma}^6} \bigg ) \,
    E_\ell (E_{\gamma_1} E_{\gamma_2})^2(1-\cos \theta_{\gamma\gamma})^2 \ ,
\label{dgamma_tlgg}
\eeqs
where
\beq
\sum_{\cal O} |\bar c_{\cal O}|^2 =
|\bar c_{\tau \ell \gamma \gamma,LR;FF}|^2 +
|\bar c_{\tau \ell \gamma \gamma,RL;FF}|^2 +
|\bar c_{\tau \ell \gamma \gamma,LR;F\tilde F}|^2 +
|\bar c_{\tau \ell \gamma \gamma,RL;F\tilde F}|^2 \ , 
\label{csqsum}
\eeq
\end{widetext}
$E_\ell$, $E_{\gamma_1}$, and $E_{\gamma_2}$ are the energies of
the daughter lepton $\ell$ and the two photons, respectively, 
and $\theta_{\gamma\gamma}$ is the angle between the 3-momenta of the
photons (i.e., $\cos\theta_{\gamma\gamma} = ({\vec k}_1 \cdot {\vec
  k}_2)/(E_{\gamma_1}E_{\gamma_2})$ ) in the $\tau$ rest frame. 
  
We note that  a two-photon final state could also arise as a radiative correction to the  decay $\tau \to \ell \gamma$,  via emission of the second photon from the initial $\tau$ or from the final-state $\ell$, where $\ell=e$ or $\mu$.  An event of this type would have an angular distribution different from that of an event in which the two photons originated directly as a consequence of the BSM physics, and the associated ${\cal L}_{eff,\tau\ell\gamma\gamma}$  in Eq. (\ref{leff_tlgg}).  Events in which a second photon is emitted as a radiative correction to a $\tau \to \ell \gamma$ decay were considered by the BABAR  experiment~\cite{aubert_babar_tlg}, were modelled by the event simulation programs used in that experiment, and were taken into account in their upper limits on 
$Br(\tau \to \ell \gamma)$.

\subsection {Study of $\tau \to \ell \gamma \gamma$ based on BABAR  limits on $\tau \to \ell \gamma$}
\label{tlggb_section}.

The BABAR experiment
searches for $\tau \to \ell \gamma$ decays \cite{aubert_babar_tlg}
were performed at the SLAC PEP-II $e^+ e^-$ storage rings,  primarily  using center-of-mass (c.m.) energy $\sqrt{s}
\simeq 10.6$ GeV at the $\Upsilon(4S)$ resonance.  
The BABAR detector is described in Ref.~\cite{BABAR}.
Charged particles were reconstructed as tracks with a 
silicon vertex tracker and a drift chamber inside a
1.5 T solenoidal magnet. A CsI(Tl) electromagnetic calorimeter
identified  electrons and photons, and a ring imaging
Cherenkov detector  identified charged
pions and kaons. The flux return of the solenoid was instrumented
with resistive plate chambers, and limited streamer
tubes were used to identify muons.

Events ascribed to
the reaction $e^+e^- \to \tau^+\tau^-$ were selected, and events of
the form $\tau^\pm \to \ell^\pm \gamma$,  were
identified  by a $\ell, \gamma$
pair with an
invariant mass and total energy in the c.m. frame close to $m_\tau= 1.777$ GeV/$c^2$ 
and $\sqrt{s}/2$, respectively. Another  $\tau^\pm$ decay in the opposite detector hemisphere was used
as a tag.
Important backgrounds arose from the
reaction $e^+e^- \to \tau^+\tau^-\gamma$ yielding a hard photon when
one  $\tau$ underwent a SM decay to an $\ell$ and a
neutrino anti-neutrino pair.  Other backgrounds for the $\tau \to \ell
\gamma$ search arose from the reaction $e^+e^- \to \ell^+\ell^-\gamma$
and from hadronic $\tau$ decays with particle mis-identification.

The signal-side hemisphere was required to  contain one photon
with c.m. energy 
$>1$ GeV, with no other photon with energy $>100$  MeV in the laboratory
frame. The signal had to contain one track identified as an electron or muon within the
calorimeter acceptance with c.m. momentum 
less than $0.77 \sqrt{s}/2$.  Muons were also required to have momentum greater than
0.7 GeV/$c$ in the laboratory frame. 
In addition, the cosine
of the opening angle between the signal track and
signal photon was required to be less than 0.786 characterizing the back-to-back distribution of $\tau \to \ell \gamma$ events in the $\tau$ rest frame. Neural net cuts were also applied to the BABAR data.

Signal decays were identified by two kinematic variables:
the energy difference $\Delta E=E_{\ell\gamma}^{c.m.}-\sqrt{s}/2$, where $E_{\ell\gamma}^{c.m.}$ is the c.m. energy of the $\ell \gamma$ pair,   and the beam energy
constrained $\tau$ mass (mEC), obtained from a kinematic fit  after requiring the c.m. $\tau$ energy to be $\sqrt{s}/2$; the origin of the $\gamma$ candidate was assigned to  the point of
closest approach of the signal lepton track to the $e^+e^-$ 
collision axis \cite{aubert_babar_tlg}.

Limits on the decays $\tau \to e \gamma\gamma$ and $\tau \to \mu \gamma\gamma$ were obtained using the results of the BABAR experiment searching for $\tau \to e \gamma$ and $\tau \to \mu \gamma$ decays.
Using Eq. (\ref{dgamma_tlgg}), we simulated $1{\times}10^7$ events for each \tl~process applying momentum and energy resolutions (smearing)  for the charged track and photons as reported by Ref.~\cite{BABAR} and applying the cuts indicated above (except for the neural net cuts) to select events.  Then, without the resolution effects applied, we constructed the mEC and $\Delta E$ variables for the \tl~events which passed the cuts and were within the BABAR detector acceptance. The mEC and $\Delta E$  variables were then smeared according to their reported  resolutions  \cite{aubert_babar_tlg}. 

\begin{figure}

\includegraphics[scale=0.36]{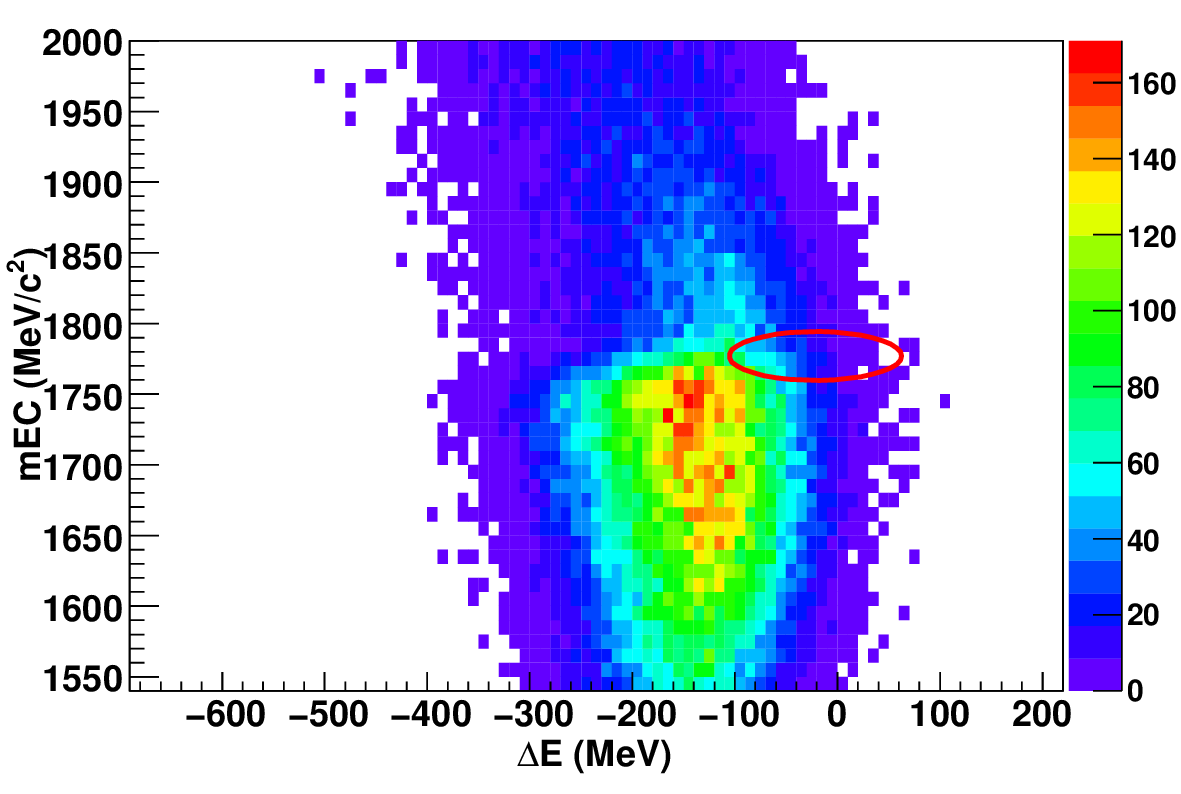}
\caption{mEC vs. $\Delta E$ for simulated $\tau \to e \gamma \gamma$ events. The red ellipse indicates the signal region where events would have passed cuts for $\tau \to e \gamma$ ~\cite{aubert_babar_tlg}.
}
\label{TEGG}
\end{figure}

Figure \ref{TEGG} shows a plot of mEC vs. $\Delta E$ for simulated $\tau \to e \gamma \gamma$ events after the cuts and resolution smearing. 
Compared with $\tau \to e \gamma$  in Ref. \cite{aubert_babar_tlg}, the plot of mEC vs. $\Delta E$ for $\tau \to e \gamma \gamma$ is widely distributed due to the requirement for the second gamma to have $E<100$ MeV if in the signal side hemisphere  or to be outside the detector acceptance.
The red ellipse in Figure \ref{TEGG} represents the signal region for ${\tau}{\to} e {\gamma}$ used by the BABAR analysis including the observed shift in position due to radiative effects~\cite{aubert_babar_tlg}. 
This elliptical region contains the simulated $\tau \to e \gamma \gamma$ events which would have been classified as consistent with the  $\tau \to e \gamma$  signal representing an efficiency  of 
$\epsilon_{e \gamma \gamma}=1.2\times 10^{-4}$ 
compared to $\epsilon_{e \gamma }=0.50$ for our simulation efficiency for $\tau \to e \gamma$.  The estimated uncertainty in the ratio $\epsilon_{e \gamma }/\epsilon_{e \gamma\gamma}$ (used below) is approximately 10\%.

To obtain the limits on $\tau\to \ell \gamma$, BABAR used the numbers of observed events and  the numbers of  the expected background events in the signal ellipse leading to  $Br(\tau \to e \gamma)<3.3 \times 10^{-8}$ and $Br(\tau \to \mu \gamma)<4.4 \times 10^{-8}$. For $\tau \to e \gamma$  ($\tau \to \mu \gamma$), 0 (2)  events were observed and the expected background  was $1.6\pm 0.4$ ($3.6\pm 0.7$), respectively. In order to avoid complications of  estimating the expected backgrounds for $\tau \to \ell \gamma$ in the presence of  \tl~ decays \cite{hearty_thanks}, we used a  conservative approach
and based the following limits on only the number of events observed by BABAR in the signal ellipses: $Br'(\tau \to e \gamma)<6.1 \times 10^{-8}$ and $Br'(\tau \to \mu \gamma)<9.1 \times 10^{-8}$.

Then, we found the limit 
\beq
Br(\tau \to e \gamma \gamma) <\frac{Br'(\tau \to e \gamma)\times \epsilon_{e \gamma}}{\epsilon_{e \gamma \gamma}}=2.5\times 10^{-4}.  
\label{br_tegg_limit}
\eeq
For the $\tau \to \mu \gamma \gamma$ case, we had  
$\epsilon_{\mu \gamma \gamma}=7.2 \times 10^{-5}$ 
and $\epsilon_{\mu \gamma}=0.46$
resulting in 
\beq
Br(\tau \to \mu \gamma \gamma) <\frac{Br'(\tau \to \mu \gamma)\times \epsilon_{\mu \gamma}}{\epsilon_{\mu \gamma \gamma}}=5.8\times 10^{-4}.
\label{br_tmgg_limit}
\eeq
The estimated uncertainty in the ratio $\epsilon_{ \mu \gamma}/\epsilon_{\mu \gamma \gamma}$ is approximately 10\%.
Concerning sensitivity to new physics, our upper
bounds (\ref{br_tegg_limit}) and (\ref{br_tmgg_limit}) probe BSM scales $\Lambda_{\tau\ell\gamma\gamma} \sim O(10^2)$ 
GeV  if  the $|\bar c_{\cal O}| \sim O(1)$. 

We note that the upper bounds 
$Br(\tau \to e \pi^0) < 0.80 \times10^{-7}$ 
and $Br(\tau \to \mu \pi^0) < 1.2 \times 10^{-7}$ from Belle \cite{miyazaki2007} and $Br(\tau \to \mu \pi^0) < 1.1 \times 10^{-7}$ from BABAR \cite{aubert2007} may also be used to obtain limits on $\tau\to \ell \gamma \gamma$. However, our evaluation of these processes led to limits on $\tau\to \ell \gamma \gamma$ that were two orders of magnitude less sensitive than those presented in Eqs. (\ref{br_tegg_limit}) and (\ref{br_tmgg_limit}).

\section{$\tau \to \ell X$ }

\label{tlx_section}

In this section we obtain new constraints on the decays $\tau \to \ell
X$ where $X$ is a  weakly
interacting neutral boson that escapes without being
detected.  The latter condition is satisfied if the lifetime $\tau_X$
is sufficiently long or if $X$ decays invisibly.  Theoretical
motivations for searching for such emission were discussed in Section
\ref{theory_motivation_section}.

The signature for the decay $\tau \to \ell X$ is 
a monochromatic peak in the energy of the daughter lepton $\ell$ in
the $\tau$ rest frame at the value
\beq
E_\ell = \frac{m_\tau^2 + m_\ell^2 - m_X^2}{2m_\tau}
\label{Eell}
\eeq
where $m_X$ is the mass of the $X$ particle.  This type of search involves
an analysis of the energy or momentum spectrum of the daughter lepton in
the $\tau$ decay.
A different approach to setting an upper limit on $Br(\tau \to e X)$
and $Br(\tau \to \mu X)$ is based on the fact that if such events occurred
and were included together with events from the corresponding SM leptonic
decays of the $\tau$, they would alter the observed rates of the respective decays.

Measurements of the individual branching ratios for 
$\tau\to \nu_\tau e \bar \nu_e$ and $\tau \to
\nu_\tau \mu \bar\nu_\mu$ have been carried out, with the results 
\cite{pdg}
\beq
Br(\tau \to \nu_\tau e \bar\nu_e) = 0.1782 \pm 0.0004 
\label{Br_tau_e}
\eeq
and
\beq
Br(\tau \to \nu_\tau \mu \bar\nu_\mu) = 0.1739 \pm 0.0004 \ . 
\label{Br_tau_mu}
\eeq

\begin{figure}
\includegraphics[scale=0.4]{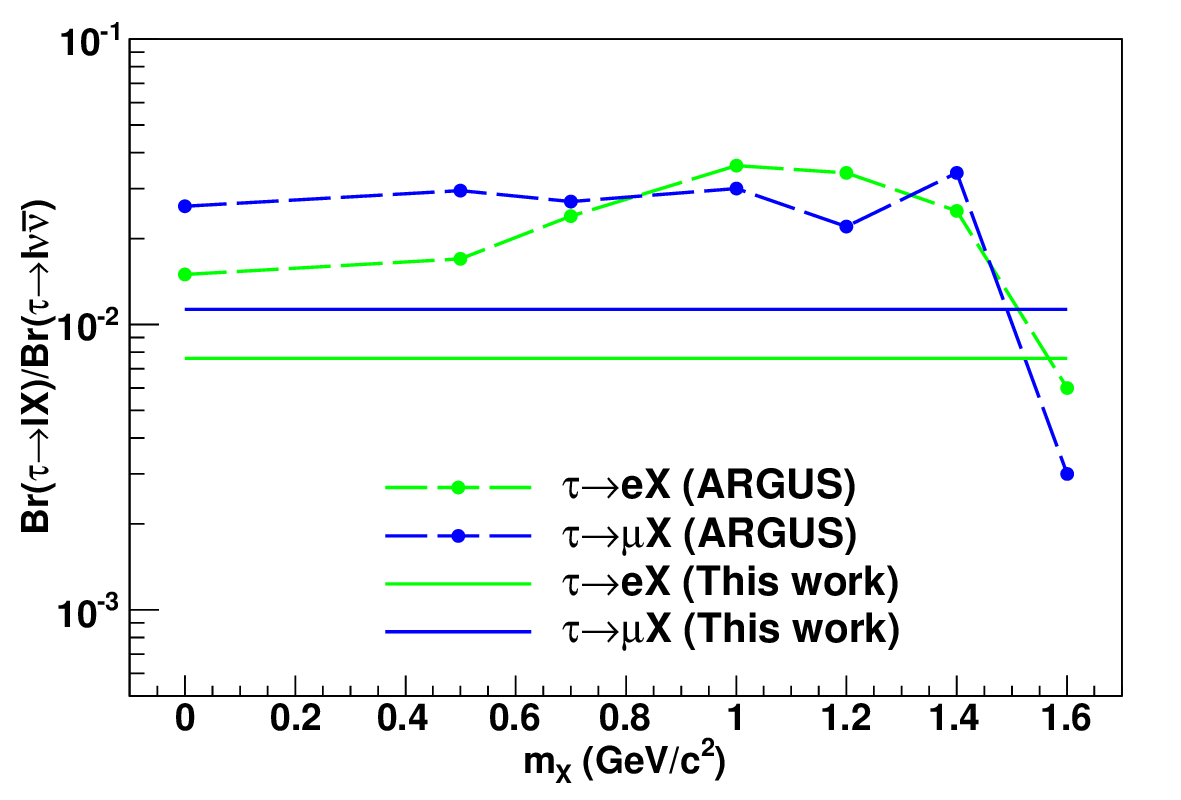}
\caption{Limits (95\% C.L.) on $Br(\tau \to \ell X)/Br(\tau \to \ell \nu_\tau \bar\nu_{\ell})$ from ARGUS \cite{albrecht95} (filled points) and this work (solid lines).
}
\label{tlx_limits_figure}
\end{figure}

The measured branching ratios
(\ref{Br_tau_e}) and (\ref{Br_tau_mu}) and the $\tau$ lifetime 
$\tau_\tau = (2.903 \pm 0.005) \times 10^{-13}$ s \cite{pdg} can  be used to obtain the decay rates to compare with SM calculations. 
Using the formulation in \cite{schael}, the calculated
values for the branching ratios (denoted by superscript $(c)$) are 
$Br^{(c)}(\tau \to \nu_\tau e \bar\nu_e) = 0.17781 \pm 0.00031$ and 
$Br^{(c)}(\tau \to \nu_\tau \mu \bar\nu_\mu)=0.17293 \pm 0.00030$. Then, the ratios of 
experimental to calculated decay rates are \cite{typo,pft}
\beq
S_{\tau\to e}=   \Gamma_{\tau\to e}/\Gamma^{(c)}_{\tau\to e,~{\rm SM}}=1.0022\pm 0.0028
\eeq
and 
\beq
S_{\tau\to \mu}=\Gamma_{\tau\to \mu}/\Gamma^{(c)}_{\tau\to \mu,~{\rm SM}}=1.0056 \pm 0.0029
\eeq
with
the following 95\% C.L. \cite{feldman_cousins} limits
\beq
S_{\tau\to e } < 1.008
\label{Staue}
\eeq
and
\beq
S_{\tau\to\mu} < 1.011 \ .
\label{Staumu}
\eeq
Eqs. (\ref{Staue}) and (\ref{Staumu}) correspond to the 95\% C.L. limits on the branching ratios of $\tau \to \ell X$ relative to $\tau \to \ell \nu \bar{\nu}$
\beq
\frac{Br(\tau\to e X)}{Br(\tau \to \nu_\tau e 
\bar\nu_e )} < 0.008
\label{StaueX}
\eeq
and
\beq
\frac{Br(\tau\to \mu X)}{Br(\tau \to \nu_\tau \mu \bar\nu_\mu )} < 0.011 \ .
\label{StaumuX}
\eeq
These limits are plotted in Fig.~\ref{tlx_limits_figure} 
along with the previous results from the ARGUS experiment \cite{albrecht95}. Using the measured $\tau\to \nu_\tau\ell \bar\nu_\ell$ branching ratios in Eqs. (\ref{Br_tau_e}) and (\ref{Br_tau_mu}), we found $Br(\tau\to e X)<1.4 \times 10^{-3}$ and $Br(\tau\to \mu X)<2.0 \times 10^{-3}$.  

Our new upper bounds (\ref{StaueX}) and (\ref{StaumuX})
yield  improved lower bounds on the weighted decay constants $F_{\tau \ell}$, $\ell=e, \ \mu$, appearing in the effective Lagrangian for $\tau \to \ell X$. For example, in the notation of Table 1 of Ref. \cite{calibbi2020}, at an illustrative mass $m_X = 0.6$ GeV, our bounds
increase the lower limit on $F_{\tau e}$ from $4.3 \times 10^6$ GeV to $\sim 7 \times 10^6$ GeV and increase the lower limit on
$F_{\tau \mu}$ from $3.3 \times 10^6$ GeV to $\sim 6 \times 10^6$ GeV. 
The limits found for $\tau \to \ell X$ decays also apply to three-body decays of the form $\tau \to \ell X X$, for which no previous bounds have been reported.

\section{Conclusions}

Using an analysis of data from searches for $\tau \to e\gamma$
and $\tau \to \mu \gamma$ performed by the BABAR experiment, we have
obtained the first upper limits on the branching ratios
$Br(\tau \to e \gamma\gamma)$ and $Br(\tau \to \mu \gamma\gamma)$.  
We have also presented improved upper limits on  $Br(\tau \to \ell X)$  where $\ell$
denotes $e$ or $\mu$ and $X$   is a weakly interacting boson with mass  $m_X<1.6$ GeV/$c^2$ that escapes
detection. We expect that these decay modes can be searched for with considerably higher sensitivity at Belle II \cite{belle2_tdr}.

\begin{acknowledgments}
 We would like to thank C. Hearty for reading the manuscript and B. Velghe for technical assistance. This work was supported by NSERC  grant no.  SAPPJ-2018-0017 (Canada); JSPS KAKENKI grant No. 19K03888 (Japan); and the  NSF Grant NSF-PHY-1915093 (U.S.).
\end{acknowledgments}


\begin{thebibliography}{99}

\bibitem{bryman_clifford}
D. A. Bryman and E. T. H. Clifford, Phys. Rev. Lett. {\bf 57}, 2787 (1986).

\bibitem{bilger}
R. Bilger et al., Phys. Lett. B {\bf 446}, 363 (1999). 

%
\bibitem{jodidio}
  A. Jodidio et al., Phys. Rev. D {\bf 34}, 1967 (1986); Err. Phys. Rev.
  D {\bf 37}, 237 (1988). .

%
\bibitem{twist}
 R. Bayes et al. (TWIST Collab.), Phys. Rev. D {\bf 91}, 052020 (2015).

%
\bibitem{triumf_muex}
  A. Aguilar-Arevalo et al. (PIENU Collab.). Phys. Rev. D {\bf 101}, 052014
  (2020) [arXiv:2002.09170].

\bibitem{eichler}
R. Eichler et al., Phys. Lett. B {\bf 175}, 101 (1986).

\bibitem{triumf88}
C. E. Picciotto et al., Phys. Rev. D {\bf 37}, 1131 (1988).

\bibitem{triumf_pilnx}
A. Aguilar-Arevalo et al. (PIENU Collab.), Phys. Rev. D {\bf 103}, 052006 (2021) [arXiv:2101.07381].

\bibitem{yamazaki}
T. Yamazaki et al., Phys. Rev. Lett. {\bf 52}, 1089 (1984).

\bibitem{baker87}
N. J. Baker et al., Phys. Rev. Lett. {\bf 59}, 2832 (1987). 


\bibitem{bnl2}
V. V. Anisomovsky et al., Phys. Rev. Lett. {\bf 93}, 031801 (2004). 

\bibitem{na62_klnx}
E. Cortina Gill et al. (NA62 Collab.), JHEP, in press; [arXiv:2011.11329] and [arXiv:2103.15389v1].

%
\bibitem{albrecht95}
 H. Albrecht et al. (ARGUS Collab.), Zeit. f. Phys. C {\bf 68}, 25 (1995).


\bibitem{pdg}
 Particle Data Group, Review of Particle Properties online at
 http://pdg.lbl.gov.

\bibitem{meg77}
 W. J. Marciano and A. I. Sanda, Phys. Lett. B {\bf 67}, 303 (1977);
 B. W. Lee, S. Pakvasa, R. E. Shrock, and H. Sugawara, Phys. Rev. Lett.
 {\bf 38}, 1230 (1977);
 S. M. Bilenky, S. T. Petcov, and B. Pontecorvo, Phys. Lett. {\bf 67}, 309
 (1977).

\bibitem{ls77}
B. W. Lee and R. E. Shrock, Phys. Rev. D {\bf 16}, 1444 (1977).

\bibitem{meg}
A. M. Baldini et al. (MEG Collab.), Eur. Phys. J. C {\bf 76}, 434 (2016).

\bibitem{bolton_lampf_megg}
R. D. Bolton et al., Phys. Rev. D {\bf 38}, 2077
(1988). See also D. Grosnick et al., Phys. Rev. Lett. {\bf 57}, 3241 (1986).

\bibitem{sindrum}
  U. Bellgardt et al. (SINDRUM Collab.), Nucl. Phys. B {\bf 299}, 1 (1988).

\bibitem{hall_suzuki}
L. J. Hall and M. Suzuki, Nucl. Phys. B {\bf 231}, 419 (1984). 

\bibitem{ihl_meg}
  I-H. Lee, Phys. Lett. B {\bf 138}, 121 (1984); Nucl. Phys. B {\bf 246}, 120
  (1984).

\bibitem{bhs}
R. Barbieri, L. J. Hall, and A. Strumia, Nucl. Phys. B {\bf 445}, 219 (1995).


\bibitem{hisano2009}
J. Hisano, M. Nagai, P. Paradisi, and Y. Shimizu, JHEP 12(2009) 030. 

\bibitem{barber_shrock}
J. Barber and R. E Shrock, Phys. Lett. B {\bf 139}, 427 (1984).

\bibitem{etc}
  S. Dimopoulos and L. Susskind, Nucl. Phys. B {\bf 155}, 23 (1979);
  E. Eichten and K. Lane, Phys. Lett. B {\bf 90}, 125 (1980).

\bibitem{ckm}
  See, e.g., T. Appelquist and J. Terning, Phys. Rev. D {\bf 50}, 2116 (1994);
  T. Appelquist and R. Shrock, Phys. Lett. B {\bf 548}, 204 (2002);
  T. Appelquist, M. Piai, and R. Shrock, Phys. Rev. D {\bf 69},
  015002 (2004); Phys. Lett. B {\bf 593}, 175 (2004); T. Appelquist,
  N. C. Christensen, M. Piai, and R. Shrock, Phys. Rev. D {\bf 70}, 093010
  (2004) and references therein.

\bibitem{tc_pdg}
  R. S. Chivukula, M. Narain, and W. J. Womersley in Ref. \cite{pdg},
  {\it op cit.}.
\bibitem{dml}
T. Appelquist, M. Piai, and R. Shrock, Phys. Lett. B {\bf 593}, 175 (2004).

\bibitem{wtc}
  B. Holdom, Phys. Lett. B {\bf 150}, 301 (1985); K. Yamawaki, M. Bando,
  and K. Matumoto, Phys. Rev. Lett. {\bf 56}, 1335 (1986); T. Appelquist,
  D. Karabali, and L. C. R. Wijewardhana, Phys. Rev. Lett. {\bf 57}, 957
  (1986).

\bibitem{qcconferences}
  See, e.g., Proceedings of SCGT-15 (Strongly Coupled Gauge
  Theories), Nagoya University,
  Int. J. Mod. Phys. A {\bf 32}, 1747007 (2017);
  Simons Workshop on Continuum and Lattice Approaches to Conformal and
  Quasiconformal Gauge Theories, Stony Brook Univ., Jan. 2018,
  http://scgp.stonybrook.edu/archives/21358;
  G. Caccapaglia, C. Pica, and F. Sannino, Phys. Rept. {\bf 877}, 1 (2020)
  and references therein.

\bibitem{quasiconformal}
Y. Aoki et al. Phys. Rev. D 89, 111502 (2014);
T. Appelquist et al., Phys. Rev. D 93, 114514 (2016);
T. A. Ryttov and R. Shrock, Phys. Rev. D {\bf 94}, 105014 (2016);
Phys. Rev. D {\bf 95}, 105004 (2017); 
Y. Aoki et al., Phys. Rev. D 96, 014508 (2017);
T. Appelquist et al., Phys. Rev. D 99, 014509 (2019);
Z. Fodor, K. Holland, J. Kuti and C. H. Wong, PoS LATTICE2018, 196 (2019);
T. Appelquist, J. Ingoldby, and M. Piai, arXiv:2012.01237 and references
therein.

\bibitem{wilczek_zee}
F. Wilczek and A. Zee, Phys. Rev. Lett,. {\bf 38}, 531 (1977). 

\bibitem{dcl}
A. Delgado, C. Garcia Cely, T. Han, and Z. Wang, Phys. Rev. D {\bf 84}, 073007 (2011);
S. Biondini, O. Panella, G. Pancheri, Y. N. Srivastava, and L. Fan\`o, Phys. Rev. D 
{\bf 85}, 095018 (2012); A. Alloul, M. Frank, B. Fuks, and M. Rausch de Traubenberg, 
Phys. Rev. D {\bf 88}, 075004 (2013); K. S. Babu, A. Patra, and S. K. Rai, 
Phys. Rev. D {\bf 88}, 055006 (2013); T. Ma, B. Zhang, and G. Cacciapaglia, 
Phys. Rev. D {\bf 89}, 093022 (2014); R. Leonardi, O. Panella, and L. Fan\`o, 
Phys. Rev. D {\bf 90}, 035001 (2014) and references therein. 
  
\bibitem{wilczek82}
F. Wilczek, Phys. Rev. Lett. {\bf 49}, 1549 (1982).

\bibitem{grinstein_preskill_wise}
  B. Grinstein, J. Preskill, and M. B. Wise, Phys. Lett. B {\bf 159}, 57
  (1985).

 
\bibitem{cmp_majoron}
  Y. Chikashige, R. N. Mohapatra, and R. D. Peccei,
  Phys. Lett. B {\bf 98}, 265 (1981).

\bibitem{gelmini_roncadelli}
G. B. Gelmini and M. Roncadelli, Phys. Lett. B {\bf 99}, 411 (1981). 

  
\bibitem{gelmini_nussinov_yanagida}  
  G. B. Gelmini, S. Nussinov, and T. Yanagida,
  Nucl. Phys. B {\bf 219}, 31 (1983). 

\bibitem{berezhiani}
Z. G. Berezhiani and M. Yu. Khlopov, Zeit. Phys. C {\bf 49}, 73 (1991).

\bibitem{feng}
  J. L. Feng, T. Moroi, H. Murayama, and E. Schapka, Phys. Rev. D {\bf 57},
  5875 (1998). 

\bibitem{jaeckel_ringwald}
  J. Jaeckel and A. Ringwald, Ann. Rev. Nucl. Part. Sci. {\bf 60}, 405 (2010).

  
\bibitem{jaeckel}
  J. Jaeckel and M. Spannowsky, Phys. Lett. B {\bf 753}, 482 (2016)

  
\bibitem{heeck_rodejohann}
J. Heeck and W. Rodejohann, Phys. Lett. B {\bf 776}, 385 (2018).

 
\bibitem{calibbi_review}
L. Calibbi and G. Signorelli, Rev. Nuovo Cim. {\bf 41} 71 (2018).


\bibitem{neubert_alp}
M. Bauer, M. Neubert, S. Renner, M. Schnubel, and A. Thamm, Phys. Rev. Lett.
{\bf 124}, 211803 (2020).

\bibitem{cornella}
C. Cornella, P. Paradisi, and O. Sumensari, JHEP {\bf 01} (2020) 158.

  
  

\bibitem{endo_belle2}
M. Endo, S. Iguro, and T. Kitahara, JHEP 06(2020) 040.


\bibitem{diluzio}
  L. Di Luzio, M. Giannotti, E. Nardi, and L. Visinelli,
Phys. Rept. {\bf 870}, 1 (2020).

  
\bibitem{calibbi2020}
L. Calibbi, D. Redigolo, R. Ziegler, and J. Zupan, arXiv:2006.04795. 

\bibitem{langacker_zprime}
P. Langacker and M. Pl\"umacher, Phys. Rev. D {\bf 62}, 013006 (2000).

\bibitem{heeck_zprime}
  J. Heeck, Phys. Lett. B {\bf 758}, 101 (2016).

\bibitem{ng}
W.-F. Chang and J. N. Ng, Phys. Rev. D {\bf 71}, 053003 (2005).
 
\bibitem{nuled}
S. Girmohanta, R. N. Mohapatra, and R. Shrock, Phys. Rev. D {\bf 103}
015021 (2021).

\bibitem{as}
  N. Arkani-Hamed and M. Schmaltz, Phys. Rev. D {\bf 61}, 033005 (2000).

\bibitem{ms}
E. A. Mirabelli and M. Schmaltz, Phys. Rev. D {\bf 61}, 113011 (2000).  

\bibitem{belle2_physics}
E. Kou et al., (Belle II Collab.), Prog. Theor. Exp. Phys. 
PTEP {\bf 2019}, 123C01 (2019) [arXiv:1808.10567]. 

\bibitem{snowmass_dm}
M. Battaglieri et al., arXiv:1707.04591.



\bibitem{aubert_babar_tlg}
 B. Aubert et al. (BABAR Collab.), Phys. Rev. Lett. {\bf 104}, 021802 (2010).


\bibitem{belle2021_tmg}
A. Abdesselam et al. (Belle Collab.), arXiv:2103.12994. 


\bibitem{BABAR}
B. Aubert et al. (BABAR Collaboration), Nucl. Instrum.
Methods Phys. Res., Sect. A {\bf479}, 1 (2002).




\bibitem{dreitlein_primakoff}
  J. Dreitlein and H. Primakoff, Phys. Rev. {\bf 126}, 375 (1962).  


\bibitem{bclm}
  J. D. Bowman, T. P. Cheng, L.-F. Li, and H. S. Matis, Phys. Rev.
  Lett. {\bf 41},  442 (1978).

  
\bibitem{depommier77}
P. Depommier et al., Phys. Rev. Lett. {\bf 39}, 1113 (1977).


\bibitem{povel77}
H. P. Povel et al., Phys. Lett B {\bf 72}, 183 (1977).

\bibitem{megg_bounds}
The upper limit inferred in \cite{bclm} was later superseded by the limit
$Br(\mu \to e \gamma\gamma) < 0.84 \times 10^{-8}$ from a direct search at
TRIUMF \cite{azuelos83} and subsequently improved to $Br(\mu \to e \gamma\gamma)
< 0.72 \times 10^{-10}$ from a direct search by the Crystal Box experiment
at LAMPF \cite{bolton88}.

\bibitem{azuelos83}
G. Azuelos et al., Phys. Rev. Lett. {\bf 51}, 164 (1983).

\bibitem{bolton88}
R. D. Bolton et al., Phys. Rev. D {\bf 38}, 2077 (1988). 

\bibitem{recentmu}
For some recent discussions of searches for CLFV $\mu$ decays, and
comparisons with the sensitivity of the $\mu N \to e N$ search, see,
e.g., \cite{dk_megg} and references therein. 


\bibitem{dk_megg} S.~Davidson, Y.~Kuno, Y.~Uesaka, and M.~Yamanaka, Phys. Rev. D {\bf 102}, 115043 (2020); S.~Davidson, arXiv:2010.00317.


\bibitem{hearty_thanks} C. Hearty, private communication.



\bibitem{miyazaki2007}
Y. Miyazaki et al. (Belle Collaboration), Phys. Lett. B {\bf 648}, 341 (2007). 

\bibitem{aubert2007}
B. Aubert et al. (BABAR Collaboration), Phys. Rev. Lett. {\bf 98}, 061803 (2007). 


\bibitem{schael}
S. Schael et al. (ALEPH Collaboration), Phys. Rept. {\bf 421}, 191 (2005). 

\bibitem{typo}
We note the following misprint corrections in the second paper of
Ref. \cite{pft}: (i) the left-hand sides of Eqs. (7.16) and (7.17)
should read $R^{(D)}_{e/\mu,{\rm SM}}$ and $R^{(D)}_{e/\mu}|_{\rm exp}$ 
respectively (without the over-bars); (ii) 
in the sentence above Eq. (9.14), the uncertainty in
$BR^{(c)}_{\tau \to e}$ should be 0.00031 and the right-hand side of
Eq. (9.14) should be $1.0022 \pm 0.0028$.


\bibitem{pft}
D. A. Bryman and R. Shrock, Phys. Rev. D {\bf 100}, 053006 (2019);
Phys. Rev. D {\bf 100}, 073011 (2019). 

\bibitem{feldman_cousins}
G. Feldman and R. Cousins, Phys. Rev. D {\bf 57}, 3873 (1998).

\bibitem{belle2_tdr}
 T. Abe et al. (Belle II Collab.), arXiv:1011.0352.


\end{thebibliography}
\end{document}